\documentclass[aps,pra,twocolumn,amssymb,amsmath,nofootinbib]{revtex4-2}
\usepackage{amsmath}
\usepackage{amssymb}
\usepackage{latexsym}
\usepackage{revsymb}
\usepackage{epsfig}
\usepackage[colorlinks]{hyperref}

\newcommand{\ket}[1]{\left | #1 \right \rangle}
\newcommand{\bra}[1]{\left \langle #1 \right |}

\newcommand{\up}{\uparrow}
\newcommand{\down}{\downarrow}
\newcommand{\uz}{\uparrow_z}
\newcommand{\ux}{\uparrow_x}
\newcommand{\ut}{\uparrow_{\theta}}
\newcommand{\dz}{\downarrow_z}
\newcommand{\dx}{\downarrow_x}
\newcommand{\dt}{\downarrow_{\theta}}
\newcommand{\beq}{\begin{equation}}
\newcommand{\eeq}{\end{equation}}
\newcommand{\ra}{\rangle}
\newcommand{\la}{\langle}

\newcommand{\hL}{{\hat L}}

\newif\ifsup 
\supfalse 

\begin{document}

\title{Angular Momentum Flows without anything carrying it}

\author{Yakir Aharonov$^{a,b,c}$, Daniel Collins$^{d}$,  Sandu Popescu$^{d}$}

\affiliation{$^a$ School of Physics and Astronomy, Tel Aviv University, Tel Aviv 6997801, Israel}
\affiliation{$^b$ Schmid College of Science and Technology, Chapman University, Orange, California 92866, USA}
\affiliation{$^c$ Institute for Quantum Studies, Chapman University, Orange, California 92866, USA}
\affiliation{$^d$ H. H. Wills Physics Laboratory, University of Bristol, Tyndall Avenue, Bristol BS8 1TL, UK}

\date{Jul 2024}

\begin{abstract}

Transfer of conserved quantities between two remote regions is generally assumed to be a rather trivial process: a flux of particles carrying the conserved quantities propagates from one region to another.  We however demonstrate a flow of angular momentum from one region to another across a region of space in which there is a vanishingly small probability of any particles (or fields) being present.  This shows that the usual view of how conservation laws work needs to be revisited.

\end{abstract}

\maketitle

\section{Introduction}

Conservation laws are some of the most important laws of physics. Stemming from basic symmetries of nature, they have been part of all physics theories, classical and quantum, relativistic and non-relativistic. At the same time, the conceptual basis of the conservation laws seemed well established long ago. Yet quantum mechanics always comes with surprises. In the present paper we analyse the way in which conserved quantities are exchanged between systems at two remote locations. Hitherto there appeared to be nothing interesting about this. For example, a flux of particles would carry angular momentum from one location to another. Here however we show that exchanges of conserved quantities could occur even across a region of space in which there is a vanishingly small probability of any particles (or fields) being present. 

The results in this paper follow from the discovery of the Dynamic Cheshire Cat effect \cite{dynamicCheshireCat}. When describing a particle, we associate to it various properties: momentum, angular momentum, spin, energy and so on.  But in a gedanken experiment \cite{cheshireCat} which seems to come directly from the pages of Alice in Wonderland, it was shown that (in a pre and post-selected setup) physical properties can be disembodied from the particles to which they belong.  This "Quantum Cheshire Cat" phenomenon was experimentally confirmed in \cite{cheshireCatExperiment,cheshireCatExperiment2,cheshireCatExperiment3,cheshireCatExperiment4}, and extended to more properties and particles in \cite{cheshireCatComplete,cheshireCatTwin,cheshireCatHunting,cheshireCatMultiProperties,cheshireCatTeleport,cheshireCatExchangeGrins,cheshireCatExchangeGrinsExperiment}. 

The original Cheshire Cat effect was considered in mostly "static" situations.  More recently however it has been extended by showing that the disembodied property has a life of its own, evolving dynamically over time, making it a Dynamic Cheshire Cat \cite{dynamicCheshireCat}.  

The original motivation that led to the discovery of this dynamic effect was
the desire of better understanding the issue of ``counterfactual" information processing - counterfactual measurements \cite{elitzurVaidmanBomb,interactionFreeMeasurementKwiat,hardyParadox}, counterfactual computation \cite{counterfactualComputation1,counterfactualComputation2,counterfactualComputation3,counterfactualComputation4,counterfactualComputation5}, counterfactual cryptography \cite{counterfactualCryptography,interactionFreeCryptography} and in particular, counterfactual communication \cite{counterfactualCommunication,counterportation,counterfactualCommunication2,counterfactualCommunication3,counterfactualCommunication4,counterfactualCommunication5,counterfactualCommunication6,counterfactualCommunication7,counterfactualCommunication8,counterfactualCommunication9,counterfactualCommunication10,counterfactualCommunication11,counterfactualCommunication12,counterfactualCommunication13,counterfactualCommunication14,counterfactualCommunication15}.  In all of these information processing occurs without the particles which carry the information ever being present in the information processing devices.  For example, in counterfactual communication a message is transmitted from Bob to Alice despite a vanishingly small probability of the particle ever being on Bob's side. That it is possible to transmit information without any physical system carrying it seems absurd, yet the counterfactual protocols seem to do precisely this.  The main idea put forward in  \cite{dynamicCheshireCat} is that there actually {\it is} an information carrier in the counterfactual effects, even though the {\it particle} that is supposed to be there is not present: the physical property that actually carries the information could be present in a ``disembodied" way, i.e. without the particle to which it belongs, in a Cheshire Cat like effect.

The implications of this Dynamic Cheshire Cat effect go however well beyond its original motivation.  The issue that interests us here is the connection with conservation laws. 

As it is well-known, due to relativistic constraints, physical quantities are not only conserved but are conserved {\it locally}. That is, the conserved quantity moves from one place to a nearby place - technically, there is a current of the conserved quantity.  One may however imagine a different mechanism, namely that a quantity is conserved by disappearing from one place while reappearing in a remote location. In classical mechanics such a {\it global} conservation mechanism would be conceivable. But in relativistic mechanics, even if in one particular frame conservation could be obeyed in such a global manner, in a different frame conservation would be violated, since the disappearance in one place would not be simultaneous with reappearance in another place. 

The relativistic requirement of local conservation is reflected also in the non-relativistic quantum mechanical limit in the well known case of the local probability conservation. Now only is the total probability of finding the particle at some location always 1, i.e.  $\int_{-\infty}^{\infty}|\Psi(x,t)|^2dx=1$ but Schrodinger's equation implies that the wavefunction changes in such a way that the probability distribution flows from one location to neighboring ones via the {\it probability current} $j=i(\Psi^*\frac{\partial \Psi}{\partial x}-\Psi\frac{\partial \Psi^*}{\partial x})$. 

Although it is rarely discussed in textbooks, (in fact we found no mention of this at all) we 
naturally expect that all quantum conserved quantities are conserved in this way. 

The Dynamic Cheshire Cat effect described in \cite{dynamicCheshireCat} brings however a dramatic twist since the information carrier quantity is a conserved quantity, angular momentum, and this flows from one location to neighbouring ones without any probability current for the particle itself.  How conservation acts in such a case is a fundamental issue.

Importantly however, in \cite{dynamicCheshireCat} the disembodied conserved quantity transfer has only been proved {\it indirectly}, by the use of weak measurements.  Here we give the first {\it direct} demonstration of the effect.

The paper is organised as follows. In section II we describe the set-up. In sections III and IV we prove our main result on angular momentum transfer from one side of a box to the wall on the other side, a transfer that takes place with infinitesimal probability of any particles transferring. In section V we demonstrate that no linear momentum is transmitted to the wall during the process, further confirming that particles did not travel there.  We end with Conclusions.

\section{Dynamic Cheshire Cat}

\begin{figure}[ht]
\includegraphics[width=8.5cm]{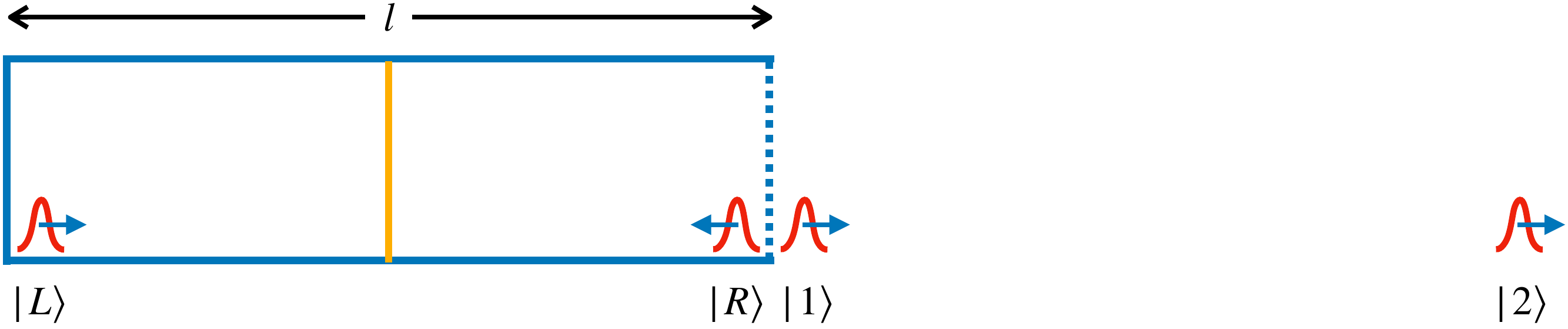}
\caption{The Dynamic Cheshire Cat.  A spin-1/2 particle starts as a wave-packet on the left side moving towards the right.  In the middle of the box is a slightly transparent and highly reflective partition.  The particle passes through the right side of the box when the spin, not shown here, is $\ket{\up_z}$, and is reflected back towards the left when the spin is $\ket{\down_z}$. }
\label{Fig:AngMomFlowFig1}
\end{figure}

We first briefly review the Dynamic Cheshire Cat \cite{dynamicCheshireCat}.  As shown in Fig. \ref{Fig:AngMomFlowFig1}, this has a spin-$1/2$ particle moving in a box which has a slightly transparent (and highly reflective) partition in the middle. The left wall of the box is completely reflective, while the wall on the right is spin-dependent.  It is fully transparent when the spin is $\ket{\up_z}$, and completely reflective when the spin is $\ket{\down_z}$.  The particle starts at the left side of the box moving towards the right in the state $\ket{L} \ket{\up_z}$. The particle will move back and forth in the box, as described below.  We shall take the mass and velocity of the particle to be large enough that the spread of the wave-packet remains small for the duration of the experiment.

At the end of the experiment we measure if the particle is in the left side of the box, where it was initially. When we find it there - which in our experiment will happen almost always - we measure its spin. It is the evolution of the spin that interests us.

In more details, we define T to be the time for the particle to travel from the starting point to the central partition, reflect off it, continue to the left wall, reflect off that and return to its starting position.  If the particle passes through the central partition, then since the right wall of the box is transparent to $\ket{\up_z}$, at time T it will be just outside the right side of the box heading to the right, in a state we call $\ket{1}$. Once outside, the particle continues to move away from the box.  This means that 
once the particle with spin $\ket{\up_z}$ passes through the mid partition into the right side of the box, it never returns to the left side.  

At time T the state has evolved as 
\beq
\ket{L} \ket{\up_z} \xrightarrow{T} ( \cos{\epsilon} \ket{L} + i \sin{\epsilon} \ket{1}) \ket{\up_z},   
\eeq
where $\epsilon=\pi/(2N)$ for some integer $N$ is a small parameter describing the transmissivity of the central partition, and the phase factor $i$ is picked up when passing through the central partition. 

After $2N$ rounds at time $2NT$, the state will have evolved as
\beq
\ket{L} \ket{\up_z} \xrightarrow{2NT} \left( \cos^{2N}{\epsilon} \ket{L} + i \sum_{k=1}^{2N} \sin{\epsilon} \cos^{2N-k}{\epsilon} \ket{k} \right) \ket{\up_z},   
\eeq
where $\ket{k}$ is a state moving to the right outside the right wall at a distance $kD$ from the left wall, where $D$ is the length of the box.  Since $\cos^{2N}{(2\pi/N)} = 1 - \mathcal{O}(1/N)$ we have 
\beq
\label{upZEvolution}
\ket{L} \ket{\up_z} \xrightarrow{2NT} \ket{L} \ket{\up_z} + \mathcal{O}(\epsilon).   
\eeq
We can make these corrections as small as desired by setting $N$ large. Thus we can make the probability of finding the particle in the left-side of the box at the end of the experiment as close to 1 as desired.  

Suppose instead we start with $\ket{\down_z}$.  The right wall now appears reflective, and we define the state of the particle at the right side of the box moving towards the left as $\ket{R}$.  This evolves as 
\beq
\begin{split}
    \ket{L}\ket{\down_z} & \xrightarrow{T} \left( \cos{\epsilon} \ket{L} + i \sin{\epsilon} \ket{R} \right) \ket{\down_z} \\
    \ket{R}\ket{\down_z} & \xrightarrow{T} \left( \cos{\epsilon} \ket{R} + i \sin{\epsilon} \ket{L} \right) \ket{\down_z}.
\end{split}
\eeq
At time $2NT$ this gives:
\beq
\label{downZEvolution}
\begin{split}
\ket{L}\ket{\down_z} & \xrightarrow{2NT} \left( \cos(2N \epsilon) \ket{L} + i \sin(2N \epsilon) \ket{R} \right) \ket{\down_z} \\ & \ = \quad -\ket{L} \ket{\down_z}.
\end{split}
\eeq
The particle has moved from the left side of the box to the right side and back again, with a phase factor of $-1$.

If we started instead with $\ket{\up_x} = \frac{1}{\sqrt{2}}(\ket{\up_z} + \ket{\down_z})$, we would see it flip to $\ket{\down_x}$.  And similarly $\ket{\down_x}$ flips to $\ket{\up_x}$.  In itself this seems ok. Indeed, the spin can change because the particle tunnels to the right-side of the box and encounters the right wall, where it undergoes a spin-dependent interaction. 

But now comes the paradox.  Suppose we start with $\ket{\up_z}$ and make the wall almost reflective (i.e. take $\epsilon$ infinitesimally small) so that apart from an infinitesimal probability the particle doesn't leave the left side of the box during the experiment.  Then we wait until time $2NT$, at which we find the particle on the left side of the box and measure the spin in the $x$ direction, $\sigma_x$, and find it $\ket{\up_x}$.  What was $\sigma_x$ at the start?  

Spin $\ket{\up_z}$ is a constant of motion so doesn't change at all. The standard view is to say that it makes no sense to ask the question about the value of $\sigma_x$; it is completely undefined when the spin is $\ket{\up_z}$.  However, in \cite{dynamicCheshireCat} it was argued that we should take the $\sigma_x$ flip seriously, hence if at the end of the experiment we measure $\sigma_x$ and find it $\ket{\up_x}$, at the beginning it should have been $\ket{\down_x}$.  This implies the spin has flipped due to the wall on the right, despite the particle never being there.  Similarly if at the end of the experiment the $\sigma_x$ measurement finds $\ket{\down_x}$, at the beginning it should have been $\ket{\up_x}$.  This surprising conclusion was supported by showing that if one performs a weak measurement of $\sigma_x$ at the start, a measurement which only disturbs the original experiment infinitesimally, then conditional on finding $\ket{\up_x}$ at time $2NT$, we would find the spin as $\ket{\down_x}$ at the start.  Whilst the measurement is weak and hence has a large uncertainly in any single experiment, it can be repeated many times to gather meaningful statistics.
  
In the next section we go beyond weak measurement arguments and look directly at the change in angular momentum of the spin-dependent wall.

\section{Angular Momentum Transfer}
\label{angularMomentumTransfer}

\begin{figure}[ht]
\includegraphics[width=8.5cm]{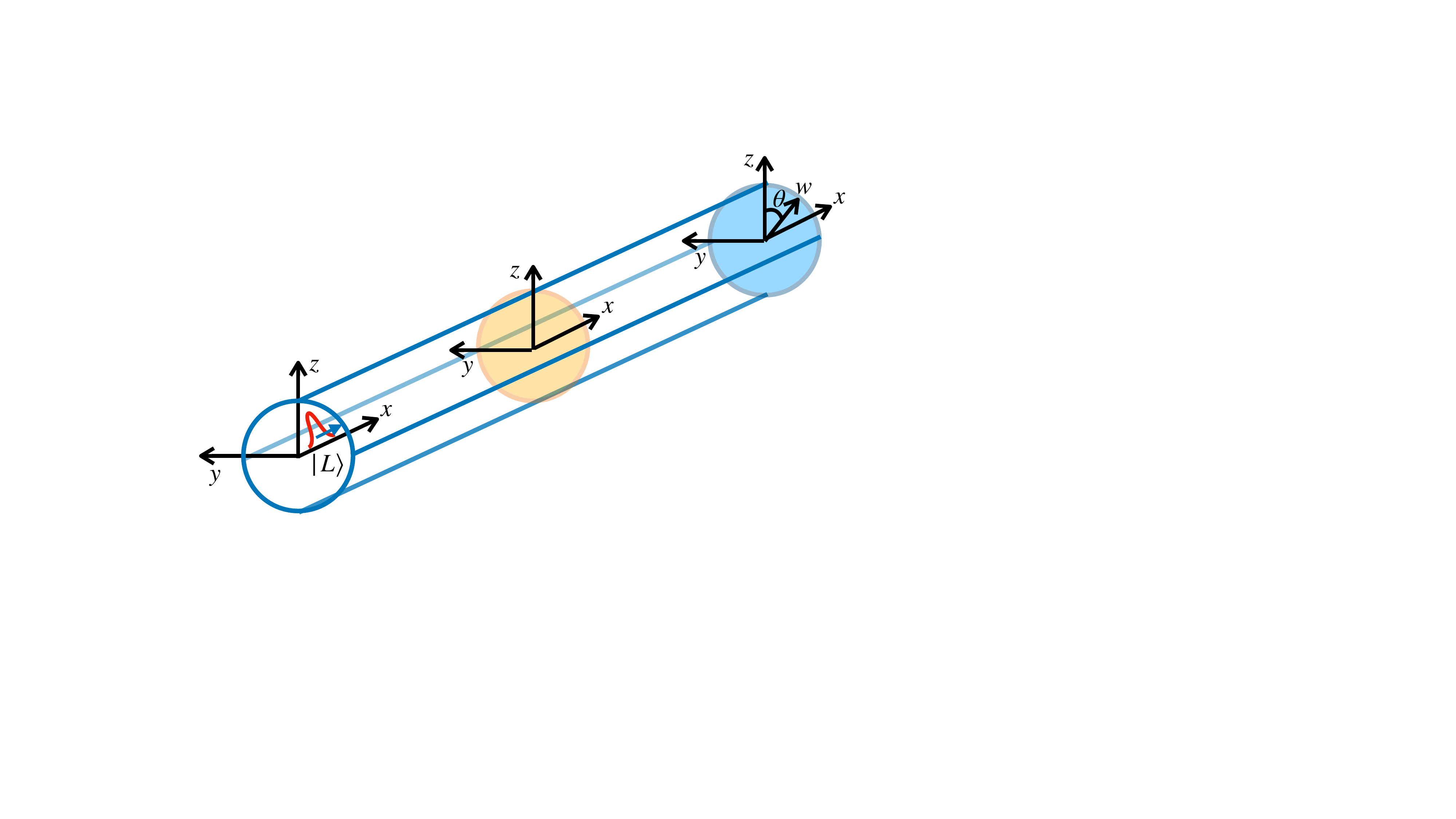}
\caption{Our thought experiment has a particle starting on the left, a highly reflective wall in the middle, and a spin-dependent wall which has a proper axis W, and which can rotate in the y-z plane, on the right.}
\label{Fig:AngMomFlowFig2}
\end{figure}

To study the angular momentum conservation including the change in angular momentum of the spin-dependent wall,  we introduce our new model, a modification of the setup of the Dynamic Cheshire Cat.   As show in Fig. \ref{Fig:AngMomFlowFig2}, the spin-dependent wall has a ``proper" axis, whose orientation is described by the unit vector ${\bf w}$ which lies in the plane of the wall.  It is fully transparent for the particle if this has the spin parallel to ${\bf w}$ and oriented ``up" and it is completely reflective if the spin is parallel to ${\bf w}$ and oriented "down".

Let the box be aligned with the $x$-axis, and the $y$ and $z$ axes perpendicular on it.  The walls and the partition are orthogonal to the $x$-axis.  The difference between our present arrangement and the one in  \cite{dynamicCheshireCat} is that we allow the spin-dependent wall to rotate.  This can be realised, for example by letting the wall have a circular form and be held by a circular frame in which it can slide.  In practice, the ``wall" can be constructed by a combination of a magnetic field and an electric potential, generated by an arrangement of a magnet and a capacitor, both of which could rotate together. We denote by $\theta$ the angle by which the wall is rotated around the $x$-axis, i.e. the angle between the wall's proper axis ${\bf w}$ and the $z$-axis.  Finally, we take its moment of inertia to be very large, so that for the duration of the experiment we can ignore its movement.  This allows us to take the free Hamiltonian of the wall to be zero.

The reason we consider a rotating wall is that we want to be able to measure changes in the $x$-component  of its angular momentum. 
In the original set-up of \cite{dynamicCheshireCat} the wall's proper axis was taken to be along the $z$-axis. This however makes the angular momentum $\hL_x$ completely undefined, so a shift of it is unobservable. (This is also a non-normalisable state, and impossible to make in practice as there will always be a small deviation from perfect $z$ in any real physical system). For this reason, we take the state of the wall to be 
\beq 
\label{spinDependentWall}
\ket{\Phi}_w=\int_{-\pi}^{\pi}\Phi(\theta)\ket{\theta}_w d\theta
\eeq
where $\Phi(\theta)$ is a wave packet with nonzero spread. 

The results below apply in fact for any $\Phi(\theta)$. We are however interested in the case when $\Phi(\theta)$ is non-zero only in the region $-\Delta \theta\leq \theta\leq \Delta \theta$ which we can take as narrow as we want, to approximate our original experiment. (If $\theta$ is large then there will be significant movement of the particle from the left-side of the box into the right-side, which spoils the original effect.)  Note that $\Delta \theta$ is independent of $\epsilon$: we can take both small independently of one another to achieve our desired behaviour.

We will be interested in the case when initially the particle is in the left-side of the box, with spin ``up" along the $z$-axis. Putting all together, the initial state is
\beq 
\label{initialState}
|L\ra|\uz\ra|\Phi\ra_w=\int_{-\pi}^{\pi}|L\ra|\uz\ra \Phi(\theta)|\theta\ra_w d\theta.
\eeq

Let now the particle evolve. Its time evolution when the wall is in direction $|\theta\ra$ is identical to that when the wall was in direction $z$ (i.e. $\theta=0$), Eqs. \eqref{upZEvolution} and \eqref{downZEvolution}, but with the spin states also rotated to the corresponding directions $\up_{\theta}$ and $\down_{\theta}$ (see Appendix \ref{AppendixAngularMomentumTransfer}). 

Suppose now that, as in the original scenario, at time $2NT$ we measure whether or not the particle is in the left side of the box and if we find it there (which happens with almost certainty), we measure its $x$-spin component. 

If we find the particle with spin $\sigma_x=+1$ the final state of the wall is (up to normalisation, $|\Phi+\ra_w$,
\begin{equation}
|\Phi+\ra_w = {\frac{1}{\sqrt{2}}}\int_{-\pi}^{\pi}e^{-i\theta}\Phi(\theta)|\theta\ra_w d\theta
\end{equation}
(see Appendix \ref{AppendixAngularMomentumTransfer}).

We now got to the main result of the paper: The average of the $x$ component of the angular momentum of the wall in the state $|\Phi+\ra_w$ is $\la {\hat L}_x\ra_+=\la {\hat L}_x\ra_0-\hbar$, where $\la{\hat L}_x\ra_0$ is the initial average angular momentum (Eq. \eqref{averageAngularMomentumEq}).  Therefore the angular momentum of the wall has changed by $-\hbar$.

This is consistent with the prediction made in \cite{dynamicCheshireCat} where $\sigma_x$ of the particle changes from $\ket{\dx}$ to $\ket{\ux}$, i.e. from $-\frac{1}{2}\hbar$ to $\frac{1}{2}\hbar$. 

Similarly, when at the end of the experiment the particle is found in the left hand side and the spin $\sigma_x=-1$, the final measurement of $\la {\hat L}_x\ra_-=\la {\hat L}_x\ra_0+\hbar$, consistent with the prediction that $\sigma_x$ of the particle changes from $\ket{\ux}$ to $\ket{\dx}$, i.e. from $\frac{1}{2}\hbar$ to $-\frac{1}{2}\hbar$. 

Comment: we note that the angular momentum transfer to the wall is independent of the initial state of the wall. This allows us to approximate as closely as we want the original setup of \cite{dynamicCheshireCat}, and assure that the perturbation that bounces off the wall can be made as small as we want, while the transfer of angular momentum to the wall remains finite in each individual case, $-\hbar$ or $+\hbar$.

\section{Angular Momentum Flux}
\label{AngularMomentumFlux}

So far we have shown that angular momentum in the $x$-direction of $\pm \hbar$ transfers from the particle to the spin-dependent wall in time $2NT$.  Now we calculate the flux of this angular momentum, i.e. how much angular momentum the spin-dependent wall gains in each period of time T.  We might try to do this by measuring the angular momentum of the wall at each time $nT$, for integer $n$ where $1 \le n \le 2N$.  However that would disturb the experiment.  Instead we consider having $2N$ walls, each one placed at the right end of the box only for the period from $(n-1)T$ to $nT$.  The flux is the momentum change on each individual wall.  In \ifsup Supplemental Material, Eq. (4)\else Appendix \ref{AppendixAngularMomentumFlux}, Eq. \eqref{fluxMthPeriod}\fi, we show that when the particle is found in the left hand side and the spin $\sigma_x=+1$, the flux is
\beq
\Delta \la {\hat L}_x\ra_n \approx - \hbar \sin \frac{(2n-1) \pi}{4N} \sin \frac{\pi}{4N}.
\eeq

It varies over half a period of a sine wave, $\sin \frac{n \pi}{2N}$, for $1 \le n \le 2N$.  The sum over all $n$ matches the result from the previous section, $-\hbar$ (see \ifsup Supplemental Material Eq. (5)\else Eq. \eqref{fluxSumsCorrectly}\fi).  The flux also matches that calculated in Eq. (39) of \cite{dynamicCheshireCat} (note that in that paper a factor $\hbar/2$ for the particle being spin-$1/2$ was omitted).  That calculation was based on weak values measured in the right side of this box: this one is direct as it's based on the angular momentum received by the spin-dependent wall over time.    

\section{Linear Momentum Transfer}
\label{LinearMomentumTransfer}

In both \cite{dynamicCheshireCat} and the present article it was argued that there is no particle travelling towards the wall, {\it only} a ``disembodied'' spin. There is however a fundamental consequence of this that was not noticed in \cite{dynamicCheshireCat}: there should be no linear momentum transfer to the wall.

When the wall is precisely oriented along $z$, it is totally transparent to the particle (which is prepared $|\up_z\ra$), so it exerts no force on it. However this orientation implies there are no observational angular momentum transfer effects, so we gave the wall an initial spread in angular direction.  Now however, the particle may sometimes collide with the wall. We have to show that the observed angular momentum transfer is not due to these collisions.  We will prove this by making the linear momentum transfer arbitrarily small whilst keeping the exchange of angular momentum unchanged at $\pm\hbar$.  

Thus far the wall was considered to be at a precise location along the $x$-axis.  In order to make a change in its linear momentum observable, we now take the position of the wall to be described by a wave packet narrowly localised around the right end of the box. We denote this as $\Psi(x_w)$, centered around $x_w=0$, where it is only non-zero for $-\Delta x_w \le x_w \le \Delta x_w$.  The initial state of the wall is then 
\begin{equation}
\ket{\Phi} \ket{\Psi} = \int_{-\pi}^{\pi}\Phi(\theta)\ket{\theta}_w d\theta \int_{-\Delta x_w}^{\Delta x_w} \Psi(x_w) \ket{x_w} d x_w.
\end{equation}

In \ifsup Supplemental Material Section I \else Appendix \ref{AppendixLinearMomentumTransfer}\fi we let the state evolve, and calculate the average linear momentum transfer.  This is given in \ifsup Supplemental Material Eq. (9)\else Eq. \eqref{averageLinearMomentumTransfer}\fi, by 
\beq
\la \hat{p_f} \ra - \la \hat{p_i} \ra = 2 p_0 \int \left| \sin( \frac{\theta}{2}) \Phi(\theta) \right|^2 d\theta + \mathcal{O}(\Delta x_w),
\eeq
where $p_0$ is the initial momentum of the particle, and $\Delta x_w$ is the uncertainty of the position of the wall. 
The change in the linear momentum of the wall will therefore go to zero when $\Delta \theta$ and $\Delta x_w$ are small, more precisely $\Delta x_w \ll \hbar / (2 N p_0)$, which we achieve by choosing $\Delta x_w$ after $N$ is chosen.  This will ensure that even after all $2N$ rounds, the wavepackets remain to a good approximation in the same phase relation as in the original unperturbed experiment (see \ifsup Supplemental Material Section II\else Appendix \ref{AppendixLinearMomentumTransfer}\fi).  Crucially, the angular momentum transfer remains $\pm\hbar$ (see \ifsup Supplemental Material Section II\else Appendix \ref{AppendixLinearMomentumTransfer}\fi). 
Thus in every experiment there is a fixed angular momentum transfer and negligible linear momentum transfer.

\section{Conclusion}

In the present paper we have given direct evidence of angular momentum transfer between two remote locations across a region of space where there is a vanishingly small probability of any particles (or fields) being present.  Although presented in the particular case of angular momentum, it seems obvious that the same phenomenon allows for disembodied transfer of arbitrary conserved quantities. Our results open a new understanding of the way conservation laws work. 

\section{Acknowledgements}
D.C. and S.P. are supported by the ERC Advanced Grant FLQuant.

\bibliography{AngMomFlow}

\medskip
\appendix

\section{Angular Momentum Transfer Calculation}
\label{AppendixAngularMomentumTransfer}

Here we calculate the change in angular momentum of the spin-dependent wall in the $x$-direction, as discussed in section \ref{angularMomentumTransfer}.  The initial state is, as given in Eq. \eqref{initialState},
\beq 
|L\ra|\uz\ra|\Phi\ra_w=\int_{-\pi}^{\pi}|L\ra|\uz\ra \Phi(\theta)|\theta\ra_w d\theta.
\eeq

The evolution of the particle was given in equations \eqref{upZEvolution} and \eqref{downZEvolution} for the case where the wall was oriented ``up'' $z$.  Since the $z$ direction was nothing special, when the wall is oriented along $\theta$ we can write similar expressions for when the spin was initially polarised $\ket{\ut}$ and $\ket{\dt}$: 
\beq
\label{new evolution}
\begin{split}
|L\ra |\ut\ra&\xrightarrow{2NT} |L\ra|\ut\ra +\mathcal{O}(\epsilon) \\
|L\ra |\dt\ra&\xrightarrow{2NT} -|L\ra|\dt\ra.
\end{split}
\eeq
 
Note that since the direction $\theta$ is in the $y$-$z$ plane the states $|\uz\ra$ and $|\dz\ra$ can be decomposed as
\beq
\begin{split}
|\uz\ra&=\cos \frac{\theta}{2} |\ut\ra+ i\sin \frac{\theta}{2} |\dt\ra\\
|\dz\ra&=i\sin \frac{\theta}{2} |\ut\ra+ \cos \frac{\theta}{2} |\dt\ra
\end{split}
\eeq

The time evolution then is
\beq
\begin{split} 
&\int_{-\pi}^{\pi}|L\ra|\uz\ra\Phi(\theta)|\theta\ra_w d\theta\\
&=\int_{-\pi}^{\pi}|L\ra\Big(\cos \frac{\theta}{2} |\ut\ra+ i\sin \frac{\theta}{2} |\dt\ra\Big)\Phi(\theta)|\theta\ra_w d\theta\\
&\xrightarrow{2NT}  \int_{-\pi}^{\pi}|L\ra\Big(\cos \frac{\theta}{2} |\ut\ra-i \sin \frac{\theta}{2} |\dt\ra\Big)\Phi(\theta)|\theta\ra_w d\theta + \mathcal{O}(\epsilon)
\end{split}
\eeq
where we have used Eq. \eqref{new evolution}.

Suppose now that at the end of this evolution we measure whether the particle is in the left side of the box or not and if we find it there we measure the spin along the $x$-axis. The spin measurement can yield $\sigma_x=+1$ or $\sigma_x=-1$. We are interested in the average angular momentum of the wall for each of these two outcomes. 

When the particle is found in the left hand side and the spin $\sigma_x=+1$, the state of the wall, $|\Phi+\ra_w$, is (up to normalisation)
\beq
\begin{split}
\label{finalStateOfWall}
&\int_{-\pi}^{\pi}\Big(\cos \frac{\theta}{2} \la \ux|\ut\ra- i\sin \frac{\theta}{2} \la \ux|\dt\ra\Big)\Phi(\theta)|\theta\ra_w d\theta\\
&={\frac{1}{\sqrt{2}}}\int_{-\pi}^{\pi}e^{-i\theta}\Phi(\theta)|\theta\ra_w d\theta
\end{split}
\eeq
where we have used the scalar products 
\beq
\begin{split}
\la \ux|\ut\ra&=\la \ux|\Big(\cos \frac{\theta}{2} |\uz\ra-i\sin \frac{\theta}{2} |\dz\ra\Big)\\
&={\frac{1}{\sqrt{2}}}\Big(\cos \frac{\theta}{2} -i\sin \frac{\theta}{2} \Big)={\frac{1}{\sqrt{2}}}e^{-i{\frac{\theta}{2}}}\\
\la \ux|\dt\ra&=\la \ux|\Big(\cos \frac{\theta}{2} |\dz\ra-i\sin \frac{\theta}{2} |\uz\ra\Big)\\
&={\frac{1}{\sqrt{2}}}\Big(\cos \frac{\theta}{2} -i\sin \frac{\theta}{2} \Big)={\frac{1}{\sqrt{2}}}e^{-i{\frac{\theta}{2}}}.
\end{split}
\eeq

The average of the $x$ component of the angular momentum of $|\Phi+\ra_w$, $\la {\hat L}_x\ra_+$  is $\la {\hat L}_x\ra_0-\hbar$ where $\la{\hat L}_x\ra_0$ is the initial average angular momentum. Indeed
\beq
\begin{split}
\label{averageAngularMomentumEq}
\la {\hat L}_x\ra_+&=\int_{-\pi}^{\pi} e^{i\theta}\Phi^*(\theta)(-i\hbar)\frac{\partial}{\partial \theta}e^{-i\theta}\Phi(\theta)d\theta\\
&=\int_{-\pi}^{\pi} \Phi^*(\theta)(-i\hbar)\frac{\partial}{\partial \theta}\Phi(\theta)d\theta-\hbar\int_{-\pi}^{\pi} \Phi^*(\theta)\Phi(\theta)d\theta\\
 &=\la {\hat L}_x\ra_0-\hbar
\end{split}
\eeq

Now to calculate the change in angular momentum we need to take the difference between the initial and final momentum.  It's tempting to simply state that the initial angular momentum of the wall along the $x$-axis is $\la {\hat L}_x\ra_0$.  However we have made a post-selection of $\bra{\up_x}$ which changed the initial angular momentum of the particle along the $x$-axis from $0$ to $-\frac{1}{2} \hbar$.  Therefore we need to check whether the initial angular momentum of the wall, conditional on the post-selection, has also changed. 

To do this we start from the final state of the particle, $\bra{\up_x}$, and final state of the wall as given in Eq. \eqref{finalStateOfWall}, and evolve the joint state backwards in time to the start.  Then we pre-select on the particle starting $\ket{\up_z}$, and calculate the initial angular momentum of the wall.  This goes as follows:
\begin{equation}
\begin{split}
&\frac{1}{\sqrt{2}} \int_{-\pi}^{\pi} e^{i\theta} \Phi^*(\theta) \bra{\theta}_w d\theta \bra{\up_x} \\
=& \frac{1}{2} \int_{-\pi}^{\pi} e^{i\theta} \Phi^*(\theta) \bra{\theta}_w d\theta \left( \bra{\up_z} + \bra{\down_z} \right) \\
=& \frac{1}{2} \int_{-\pi}^{\pi} e^{i\theta} \Phi^*(\theta) \bra{\theta}_w d\theta e^{-i \frac{\theta}{2}} \left( \bra{\up_{\theta}} + \bra{\down_{\theta}} \right) \\
\xrightarrow{2nT}& \frac{1}{2} \int_{-\pi}^{\pi} e^{i\frac{\theta}{2}} \Phi^*(\theta) \bra{\theta}_w d\theta \left( \bra{\up_{\theta}} - \bra{\down_{\theta}} \right) \\
=& \frac{1}{2} \int_{-\pi}^{\pi} e^{i \frac{\theta}{2}} \Phi^*(\theta) \bra{\theta}_w d\theta e^{-i \frac{\theta}{2}} \left( \bra{\up_z} - \bra{\down_z} \right) \\
\rightarrow& \frac{1}{2} \int_{-\pi}^{\pi} \Phi^*(\theta) \bra{\theta}_w d\theta \text{ under pre-selection on $\ket{\up_z}$,}  
\end{split}
\end{equation}
i.e. the post-selection has not changed the original state of the wall.

Thus the initial angular momentum of the wall conditional on the post-selection is still $\la {\hat L}_x\ra_0$, and the change in the angular momentum along the $x$-axis of the wall is $-\hbar$. 

\ifsup
\else

\section{Angular Momentum Flux Calculation}
\label{AppendixAngularMomentumFlux}

Here we calculate the flux in the angular momentum from the particle to the spin-dependent wall, as discussed in section \ref{AngularMomentumFlux}.  

We consider having $2N$ walls, with each one placed at the right end of the box only for the period from $(n-1)T$ to $nT$.  Then we can see the flux by calculating the momentum change on each individual wall.  We place the $n^{th}$ wall in the distribution $\Phi(\theta)$ (Eq. \eqref{spinDependentWall}).  To simplify our calculation of the $n^{th}$ term we approximate by taking all the other walls to be placed exactly in the $z$-direction.  We evolve the system up to time $2NT$, apply the post-selection, then calculate the angular momentum change of the $n^{th}$ wall.  

Since the initial wall is placed in the $z$-direction, up to time $(n-1)T$ we have:
\beq
\ket{L}\ket{\up_{z}} \xrightarrow{(n-1)T} \ket{L}\ket{\up_{z}},
\eeq
up to terms of $\mathcal{O}(\epsilon)$ which represent wave-packets outside the box, which will not be post-selected and so will not affect the calculation.  This post-selection on $\ket{L}$ will succeed with probability as close to $1$ as desired, by taking $\epsilon$ small and $\Phi(\theta)$ to be a narrow distribution around $\theta = 0$.  

Next we put the $n^{th}$ wall for one period, followed by a wall in the $z$-direction for the remaining periods, which gives
\begin{widetext}
\beq
\begin{split}
&\ket{L} \ket{\up_z} \int_{-\pi}^{\pi} \Phi(\theta) \ket{\theta} d\theta \\ =&\int_{-\pi}^{\pi} \ket{L} \left( \cos \frac{\theta}{2} \ket{\ut} + i \sin \frac{\theta}{2} \ket{\dt} \right) \Phi(\theta) \ket{\theta}_w d\theta \\
\xrightarrow{nT}& \int_{-\pi}^{\pi} \left( \ket{L} \cos \frac{\theta}{2} \ket{\ut} + i \sin \frac{\theta}{2} \left( \cos \epsilon \ket{L} + i \sin \epsilon \ket{R} \right) \ket{\dt} \right) \Phi(\theta) \ket{\theta}_w d\theta \\
=& \int_{-\pi}^{\pi} \left( \ket{L} \cos \frac{\theta}{2} \left( \cos \frac{\theta}{2} \ket{\up_z} - i \sin \frac{\theta}{2} \ket{\down_z} \right) + i \sin \frac{\theta}{2} \left(\cos \epsilon \ket{L} + i \sin \epsilon \ket{R} \right) \left(\cos \frac{\theta}{2} \ket{\down_z} -i \sin \frac{\theta}{2} \ket{\up_z} \right) \right) \Phi(\theta) \ket{\theta}_w d\theta.
\end{split}
\eeq

Then we evolve this for $(2N-n)$ periods to time $2NT$ using a spin-dependent wall placed in the $z$-direction.  We'll ignore the term in $\sin \epsilon \sin^2 \frac{\theta}{2} \ket{R} \ket{\up_z}$ as it's $\mathcal{O}(\theta^2)$, and $\theta$ is small.  Evolving using equations \eqref{upZEvolution} and \eqref{downZEvolution} for time $(2N-n)T$ instead of $2NT$, and then post-selecting on $\ket{L} \ket{\up_x}$ gives:

\beq
\begin{split}
\xrightarrow{2NT}& \int_{-\pi}^{\pi} \left( \cos^2 \frac{\theta}{2} + \sin^2 \frac{\theta}{2} \cos \epsilon + i \sin \frac{\theta}{2} \cos \frac{\theta}{2} (\cos((2N-n+1)\epsilon) - \cos((2N-n)\epsilon) \right) \Phi(\theta) \ket{\theta}_w d\theta \\
& \approx \int_{-\pi}^{\pi} \left( 1 - i \sin \theta \sin \frac{(2n-1) \pi}{4N} \sin \frac{\pi}{4N} \right) \Phi(\theta) \ket{\theta}_w d\theta \text{, taking $\cos \epsilon \approx 1$}\\
& \approx \int_{-\pi}^{\pi} \exp \left( - i \theta \sin \frac{(2n-1) \pi}{4N} \sin \frac{\pi}{4N} \right) \Phi(\theta) \ket{\theta}_w d\theta \text{, taking $e^x \approx 1 + x$.}
\end{split}
\eeq

\end{widetext}

Note that the post-selection on $\ket{\up_x}$ will succeed with probability as close to $1/2$ as desired: if we instead get $\ket{\down_x}$ the flux will have the opposite sign.

To calculate the angular momentum, note that the exponential looks like $e^{ - i \theta f(n)}$, for a particular function $f(n)$.  Comparing this with equations \eqref{finalStateOfWall} and \eqref{averageAngularMomentumEq} shows that the angular momentum change of the wall for the $n^{th}$ period is $- \hbar f(n)$, i.e.
\beq
\label{fluxMthPeriod}
\Delta \la {\hat L}_x\ra_n \approx - \hbar \sin \frac{(2n-1) \pi}{4N} \sin \frac{\pi}{4N}.
\eeq

Finally we check that the sum over all $m$ of this angular momentum change sums to the result from section \ref{angularMomentumTransfer}, $-\hbar$.
\beq
\label{fluxSumsCorrectly}
\begin{split}
\sum_{n=1}^{2N} \Delta \la {\hat L}_x\ra_n 
&= -\hbar \sin \frac{\pi}{4N} \sum_{n=1}^{2N} \sin \frac{(2n-1) \pi}{4N} \\
&= -\hbar \sin \frac{\pi}{4N} \csc \frac{\pi}{4N} \\
&= - \hbar. 
\end{split}
\eeq

\section{Linear Momentum Transfer Calculation}
\label{AppendixLinearMomentumTransfer}

In Section \ref{LinearMomentumTransfer} we discussed checking whether the particle bounces off the spin-dependent wall by checking whether the linear momentum of the wall changes.  Here we perform those calculations.

Suppose first that the axis of the wall is precisely in the $z$-direction, and the wall is prepared in a wave packed narrowly localised around the right end of the box, $\Psi(x_w)$ centered around $x_w=0$.  The evolution by the end of the experiment is given by
\beq 
\label{uZLinear}
\ket{L} \ket{\uz} \ket{\Psi} \xrightarrow{2NT} \ket{L} \ket{\uz} \ket{\Psi} +\mathcal{O}(\epsilon)
\eeq
where the $\mathcal{O}(\epsilon)$ terms contain wave packets that went through the wall. Indeed, since in this case nothing reflects back from the wall, the evolution is insensitive to the position of the wall, hence it is the same as in Eq. \eqref{upZEvolution}. Crucially, the wall received no linear momentum.

However to calculate what happens to the angular momentum of the particle around the $x$-axis, we also need to calculate what happens when the spin is polarised $\ket{\dz}$ and the axis of the wall is along the $z$-axis.  We assume the mass of the wall is much greater than the mass of the particle, so that when the particle bounces off the wall its velocity flips from $+v$ to $-v$.  Then if the wall is at $x_w$, the particle has to travel $2x_w$ further compared to $x_w=0$, so at time $t=T$ (the time for a particle to travel the length of the box) we have 
\beq
\label{dZLinear}
\ket{L} \ket{\dz} \ket{x_w} 
\xrightarrow{T} \left( \cos \epsilon \ket{L} + i e^{-i p_0 2 x_w / \hbar} \sin \epsilon \ket{R} \right) \ket{\dz} \ket{x_w},
\eeq
where $\epsilon = \pi / 2N$, $p_0 = v/m$ is the initial momentum of the particle, and $m$ is the mass of the particle.  The momentum transfer from the particle to the wall is defined by the phase change $e^{-i p_0 2 x_w / \hbar}$.  We want this phase to be small so that we don't break the interference between $\ket{L}$ and $\ket{R}$ when they recombine at the mid-box partition, and so that we don't break the coherence between the $\ket{\uz}$ and $\ket{\dz}$ states.  It needs to remain small even after $t=2 N T$ when the particle will, assuming a perfectly localized mirror and starting $\ket{\dz}$, have moved from $\ket{L}$ to $\ket{R}$ and back again.  We can achieve this for any fixed $N$ by choosing $\Psi(x_w)$ with $\Delta x_w$ such that $2 N p_0 \Delta x_w / \hbar \ll 1$.  Note that this implies that the wall has a large spread in momentum, so that when the wave function of the wall written in terms of the wall momentum, $\Psi(p_w)$, is moved by the momentum change of the particle, it still overlaps almost perfectly with the unshifted $\Psi(p_w)$.  As a result the final state of the particle and the wall is identical, up to corrections of $\Delta x_w$ that we can make as small as we want, to that obtained when the wall had a fixed location $x_w=0$.  Thus the conclusions in \cite{dynamicCheshireCat} about angular momentum remain unchanged, and there is no linear momentum transfer since the particle starts $\ket{\uz}$.

Consider now the wall has both a spread in $\theta$  and a spread in $x_w$ so that both the linear and angular momentum transfers are observable.  
The initial state is thus
\beq
\begin{split}
& \ket{L} \ket{\uz} \int \Phi(\theta) \ket{\theta} d\theta \ket{\Psi} \\ 
&= \ket{L} \int \left( \cos \frac{\theta}{2} \ket{\ut} + i \sin \frac{\theta}{2} \ket{\dt} \right) \Phi(\theta) \ket{\theta} d\theta \ket{\Psi}.
\end{split}
\eeq

We already handled the evolution of the two terms in this state in equations \eqref{uZLinear} and \eqref{dZLinear}, with $z$ instead of $\theta$.  So by taking $\Delta x_w$ small we preserve the coherence and angular momentum transfer.  Since only the second term has any linear momentum transfer, the average linear momentum transfer is given by 
\beq
\label{averageLinearMomentumTransfer}
\la \hat{p_f} \ra - \la \hat{p_i} \ra = 2 p_0 \int \left| \sin( \frac{\theta}{2}) \Phi(\theta) \right|^2 d\theta + \mathcal{O}(\Delta x_w),
\eeq
where $p_f$ is the final momentum of the wall and $p_i$ the initial.

So by choosing $\Phi(\theta)$ to be centered closely around $\theta = 0$ (i.e. by choosing $\Delta \theta$ to be small), and also taking $\Delta x_w$ small, we can make the linear momentum transfer as small as desired.  Recall that the angular momentum transfer is the same in every experiment even as $\Delta \theta \rightarrow 0$.  Only the linear momentum goes to 0.  Thus we have shown that there is a fixed angular momentum transfer and negligible linear momentum transfer, without the particle ever touching the wall.

\fi


\end{document}